\documentclass[
 ,final            
 ,numberedheadings 
  ]
  {aipproc}

\layoutstyle{6x9}

\newcommand{\beq}{\begin{equation*}}
\newcommand{\eeq}{\end{equation*}}
\newcommand{\be}{\begin{equation}}
\newcommand{\ee}{\end{equation}}
\newcommand{\beqa}{\begin{eqnarray}}
\newcommand{\eeqa}{\end{eqnarray}}
\newcommand{\bea}{\begin{eqnarray}}
\newcommand{\eea}{\end{eqnarray}}
\newcommand{\bra}{\langle}
\newcommand{\ket}{\rangle}

\newcommand{\s}{\sigma}
\newcommand{\w}{\omega}

\newcommand{\half}{\frac{1}{2}}
\newcommand{\emdash}{\hspace{1pt}---\hspace{1pt}}

\begin{document}

\title[QMC and the nature of dense matter: written in the stars?]{QMC and the nature of dense matter:\\ written in the stars?}



\author{J.~D.~Carroll}{ address={Centre for the Subatomic Structure of
    Matter (CSSM), Department of Physics, University of Adelaide, SA
    5005, Australia} }

\begin{abstract}
We discuss the recent progress in calculating the properties of `{\it
  hybrid stars}' (stellar objects similar to neutron stars, classified
by the incorporation of non-nucleonic degrees of freedom, including
but not limited to hyperons and/or a quark-matter core) using the
octet-baryon Quark-Meson Coupling (QMC) model. The version of QMC used
is a recent improvement which includes the in-medium modification of
the quark-quark hyperfine interaction.
\end{abstract}

\maketitle


\section{INTRODUCTION}

  The study of the QCD phase diagram is of great interest to the
  scientific community. Lattice QCD produces simulations which provide
  an insight into the \mbox{$\{T>0, \mu\simeq 0\}$} region of the
  phase diagram, whilst properties of QCD at finite chemical potential
  can be calculated using various models for dense hadronic matter (at
  $T\ge 0, \mu \gg 0$). In order to qualify the success of these
  models we require a method to test the predictions via experiment
  and observation. For finite chemical potentials, this requires tests
  at various
  relevant density scales. For low-density predictions we can compare
  calculations to experiments involving finite nuclei, but no
  experiment (to date) can probe the entire range of extreme densities
  believed to exist within neutron stars\footnote{In
    Ref.~\cite{QMC2007} the authors note that heavy-ion collisions may
    be able to provide insight into the properties of matter up the
    density range of 2--3$\rho_0$.}. We therefore turn our attention
  to proxy measurements which may support or refute our predictions,
  such as the masses and radii of pulsars.\par

  We focus our attention on a particular model\emdash the Quark-Meson
  Coupling model (QMC), which we shall describe in
  Section~\ref{sec:QMC} \emdash that has seen much success at each of
  these density scales, and focus in particular on the high-density
  Equation of State (EoS) for infinite matter (see
  Section~\ref{sec:infinite}). In order to examine the effects of
  quark degrees of freedom at extremely high densities we model a
  phase transition between this hadronic EoS and a quark-matter EoS as
  described by several models (see Section~\ref{sec:phasetrans}).  We
  examine the predictions for neutron star properties that arise from
  calculations involving the QMC EoS and the effect that the phase
  transition has on these properties in Section~\ref{sec:stars}.

\section{QMC}\label{sec:QMC}

  The Quark-Meson Coupling (QMC) model~\cite{QMC2008} describes
  baryons as `{bags}' of three confined quarks for which the
  energy-density is greater than that of the surrounding
  non-perturbative vacuum by $B \sim (180~{\rm MeV})^4$. These quarks
  are immersed in a mean-field of mesons which contribute to
  attractive scalar and repulsive vector potentials, which describe
  interactions in a similar way to Quantum Hadrodynamics (QHD). This
  model has seen much success in predicting the properties of finite
  nuclei, in particular hypernuclei (containing strange quarks) with
  the recent addition of a $\Lambda$--$\Sigma$ hyperfine splitting
  contribution~\cite{QMC2008}.\par

  We model baryons influenced by scalar-isoscalar $\s$,
  vector-isoscalar $\w$, and vector-isovector $\rho$ mesons, noting
  that the mean-field pseudoscalar-isovector $\pi$ field contribution
  is zero due to parity considerations. We additionally include
  leptons ($\ell \in\{e^-,\mu^-\}$) interacting with baryons in
  beta-equilibrium in our calculations.\par

  In QMC, the effective mass for the baryons has a quadratic
  form\emdash incorporating the scalar polarizability $d$ which
  self-consistently allows feedback of the scalar field\emdash given
  by
  \be \label{eq:QMC_mstar}
  M^*_B = M_B + \Sigma_B^s = M_B - w_B^s g_{\sigma N} \bra\sigma\ket +
  \frac{d}{2} \tilde{w}_B^s \left(g_{\sigma N} \bra\sigma\ket\right)^2,
  \ee
  where the factors $w_B^s$ and $\tilde{w}_B^s$ (the origins and
  values of which are discussed in further detail in
  Ref.~{\cite{QMC2008}}) account for the SU(6)-style coupling ratios
  for the various octet baryons
  $B\in\{N^{+,0},\Lambda,\Sigma^{\pm,0},\Xi^{-,0}\}$. 
  The inclusion of hyperons into these calculations does not result in
  negative effective baryon masses\emdash the occurrence of which is a
  shortcoming of QHD attributed to the linear effective mass, and the
  reason that we do not model hyperons in QHD\emdash since the scalar
  potential responds to the strength of the scalar field in a way that
  prevents this. 
  Indeed, the effective masses of all the baryons remain positive for
  any value of $\bra\s\ket$ in QMC.\par

  The couplings of the nucleons to the mesons in QMC ($g_{N \s}, g_{N
    \w}, g_{\rho}$) are fitted such that the saturation properties of
  nuclear matter are reproduced in isospin symmetric nuclear
  matter. The nucleon-meson couplings and quantities that we reproduce
  are given in Table~\ref{tab:saturation}. The couplings of the
  hyperons to the isoscalar mesons are related to these nucleon-meson
  couplings via SU(6) spin-flavor relations, 
  %
  and while the couplings of the hyperons to the isovector meson are
  unified into a single value, each vertex term includes an isospin
  factor $I_{zB}$ which splits the vector potentials within an isospin
  group.\par
  \begin{table}[!b]
    \begin{tabular}{ccc|ccc}
      \hline
                {$g_{\s N}$}  & 
                {$g_{\w N}$}  & 
                {$g_{\rho}$}  &
                {$({\cal E}/\rho_0 - \sum_{B}\rho_B M_B)$}    & 
                {$\rho_0$} &
                {$a_4$} \\
      \hline
      8.278 & 8.417 & 8.333 &
      -15.86~{MeV}        & 
      0.16~{fm$^{-3}$}  &
      32.5~{MeV} \\
      \hline
    \end{tabular}
    \caption{Meson-nucleon couplings and reproduced saturation
      properties of nuclear matter: energy per baryon; baryon density;
      and symmetry energy at saturation. The couplings are fitted to
      best reproduce the experimental values.
      \label{tab:saturation}}
  \end{table}
  The hyperon Fermi momenta (and hence, densities) are determined via
  relations between the baryon chemical potentials, according to 
  \be \label{eq:chempot}
  \mu_{i} = B_i \mu_n - Q_i \mu_e = \sqrt{k_{F_i}^2+(M_i+\Sigma_i^s)^2}+\Sigma_i^0,
  \ee
  where $B_i = +1\  \forall\  i\in B$ and $Q_i$ are respectively the
  baryon and electric charges (the latter in units of the proton
  charge) associated with the neutron and electron independent
  chemical potentials $\mu_n$ and $\mu_e$; and $\Sigma_i^{s,0}$ are
  the scalar and temporal components of the baryon self-energy.\par
  
  In QHD, the baryon self-energies at Hartree level are calculated by
  considering the tadpole diagrams as modifications to the baryon
  propagator. Following the derivation of Ref.~\cite{Serot:1984ey},
  the scalar component of the baryon self-energy is given by
  \be
  \label{eq:QHD_HartreeSelfEnergy_s} \Sigma_{B\ {\rm QHD}}^s = -
  g_{\s B}\bra\s\ket \ = \ - g_{\s B}\sum_{B'}\frac{g_{\s B'}}{m_\s^2}
  \frac{(2J_{B'}+1)}{(2\pi^3)}\int \Theta(k_{F_{B'}}-|\vec{k}|)
  \frac{M^*_{B'}}{E^*_{B'}(k)}\, d^3 k, 
  \ee
  %
  %
  where
  %
  $E^*_{B}(k) = \left(k^2+(M^*_{B})^2\right)^{1/2},$
  %
  and where $J_i = \half \ \forall \ i\in B$ is the spin of baryon
  $i$. In QMC, the scalar self-energy is modified to obtain 
  %
  Eq.~(\ref{eq:QMC_mstar}),
  which requires care to be taken with the additional
  self-consistency.\par

  Using these relations and fitting the couplings to saturation
  properties, the only independent quantities that still require
  definition are the proton and neutron Fermi momenta. These are
  constrained by requiring that the conserved total baryon density is
  equal to a selected value, and that the total electric charge of the
  system is zero. These conditions are given by
  \be
  \rho_{\rm total} = \sum_B\rho_B, \quad 0 = \sum_BQ_B\rho_B +
  \sum_\ell Q_\ell\rho_\ell.
  \ee

\section{INFINITE BARYONIC MATTER}\label{sec:infinite}

  Once the required self-consistent equations of the previous section
  have been satisfied, we can calculate the properties of dense
  matter. The species fractions $Y_i$ of various particles at a wide
  range of densities can be calculated once all of the relevant Fermi
  momenta are determined, given that
  \be
  Y_i = \rho_i / \rho,\quad 
  \rho_i = \frac{(2J_i+1)}{(2\pi)^3}\int \Theta(k_{F_i}-|\vec{k}|)
  d^3k = \frac{6k_{F_i}^3}{3\pi^2}. 
  \ee
  %
  The energy-density and pressure in QMC are analytically identical
  to those of QHD (as given in Ref.~\cite{Serot:1984ey}), noting of
  course that now the effective mass has the quadratic form of
  Eq.~(\ref{eq:QMC_mstar}).
  %
  %

  We have assumed up to this point that at extremely high densities
  ($\rho > 3 \rho_0$) hyperons will remain the correct degrees of
  freedom, however it is possible that at some density the concept of
  distinct baryons becomes inaccurate, and that the component quarks
  can become deconfined. In this case, we must consider the
  possibility of a phase transition to quark matter.\par
%
  
\section{PHASE TRANSITIONS}\label{sec:phasetrans}

  The Gibbs conditions for a first-order phase transition between two
  phases ($i,j$) are that at the phase transition point, the following
  equilibria are achieved:
  \begin{description}
    \item[Thermal Equilibrium:]The total temperature of each phase must
      be equal, thus $T_i = T_j$,
    \item[Mechanical Equilibrium:]The total pressure of each phase must
      be equal, thus $P_i = P_j$,
    \item[Chemical Equilibrium:]The independent chemical potentials of
      each phase must be equal, thus $\{\mu_\alpha\}_i =
      \{\mu_\alpha\}_j$.
  \end{description}
  %
  %
  It is possible that there exists a range of densities at which
  these conditions are met (and thus a mixed-phase exists), outside of
  which the phase with the greater pressure will be energetically
  favoured. Further details can be found in Refs.~\cite{JDC2009,JDCthesis}\par

  We calculate the baryon phase EoS at $T=0$ at increasing densities
  (using the independent chemical potentials as inputs to the
  quark-matter EoS) until we reach a point at which the pressure in
  each phase is equal (if it exists). This signals the onset of a
  mixed phase which we parameterize with the quark fraction
  $0\le\chi\le 1$ up to the phase transition to pure quark-matter at
  $\chi=1$. We then continue to calculate the quark-matter EoS with
  increasing density.



  We use two quark-matter models; the MIT bag
  model~\cite{Chodos:1974je} and the Nambu--Jona-Lasinio (NJL)
  model~\cite{Nambu:1961tp}. The former models constant current-quark
  masses, while the latter models the effect of dynamic chiral
  symmetry breaking, producing constituent-quark masses at low
  densities, and current-quark masses at large densities.\par
\begin{figure}[!b]
    \centering
    \includegraphics[angle=90,width=0.8\textwidth]{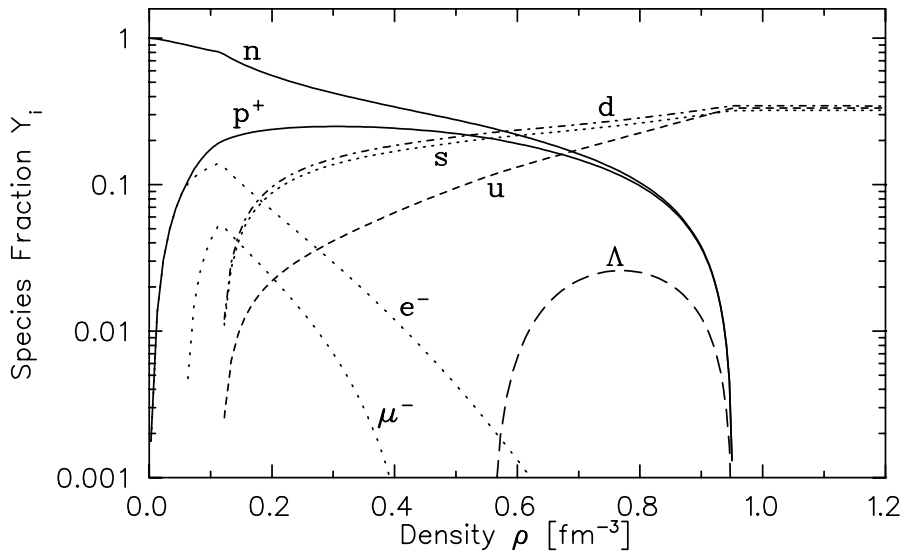}
    \caption{Species fractions $Y_i$ for hyperonic QMC with a phase
      transition to a MIT bag model quark matter via a mixed phase.
      We note that with these parameters, the transition to a mixed
      phase occurs below saturation density, $\rho_0 = 0.16~{\rm
        fm}^{-3}$.  \protect\label{fig:SpecFrac_PTQMC}}
  \end{figure}

  %
  %
  The species fractions for a hybrid EoS involving a hyperonic QMC
  baryon phase and an MIT bag model (using $B^{1/4}=180~{\rm MeV}$)
  quark phase are shown in Fig.~\ref{fig:SpecFrac_PTQMC} where we note
  the presence of baryon-, mixed-, and quark-phases.\par
  If we calculate the hybrid EoS using hyperonic QMC and NJL
  we find that the quark pressure remains lower than the baryon
  pressure for any density. This is a result of the large quark masses
  at low density and the relation between the pressure and Fermi
  momentum (via $\mu_q$). As a consequence, we find that no phase
  transition from hyperonic QMC to NJL quark-matter is possible for
  any reasonable value of $B$. We do however find that the transition
  is possible from {\it nucleonic} QMC to NJL quark-matter, in which
  case the baryon phase is not significantly softened by the inclusion
  of hyperons.\par
  %
  %
  %

\section{STELLAR SOLUTIONS}\label{sec:stars}

  In order to test these models against observations, we calculate
  properties of hybrid stars such as the total stellar mass and
  radius. The masses of these objects are comparatively easy to
  measure in the form of (particularly, binary) pulsars. The radius
  however has proven to be a greater challenge to measure.\par

  In order to calculate the predicted properties of hybrid stars we
  use the EoS (which may include a mixed- or quark-phase) as input to
  the Tolman-Oppenheimer-Volkoff (TOV) equation
  \be \label{eq:TOVdpdr}
  \frac{dP}{dR} = 
  -\frac{G\left(P + {\cal E}\right)\left(M(R)+4\pi
    R^3P\right)}{R(R-2GM(R))},\quad   M(R) = \int_0^R 4\pi r^2 {\cal E} \; dr,
  \ee
  %
  %
  %
  where ${\cal E}$ and $P$ are the energy density and pressure in the EoS,
  respectively.\par
  
\begin{figure}[!t]
    \centering
    \includegraphics[angle=90,width=0.8\textwidth]{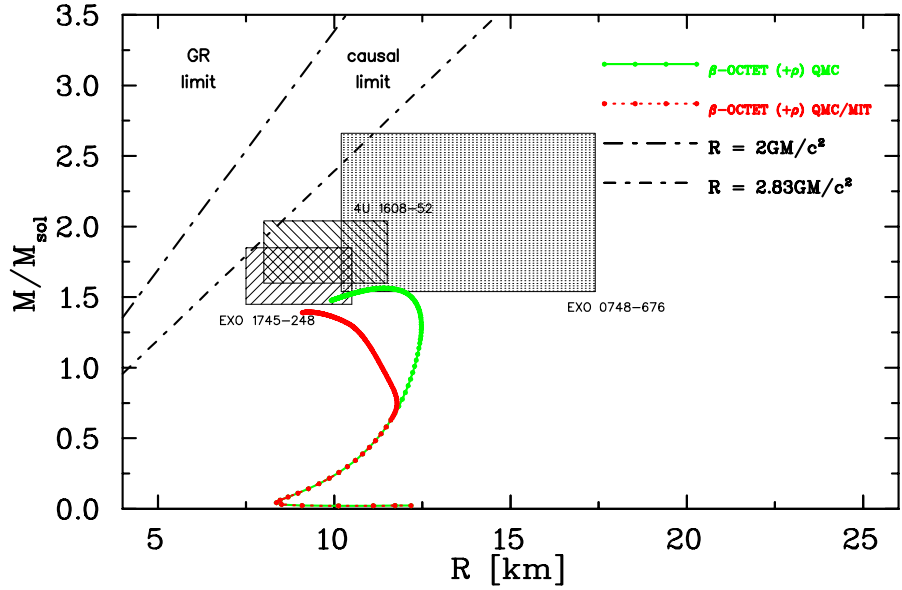}
    \caption{(Color online) Predicted total stellar mass and radius
      for hyperonic QMC with and without a phase transition to quark
      matter modelled with the MIT bag model Also shown are the data
      points from Refs.~\cite{Ozel:2006bv} (EXO 0748-676),
      \cite{Guver:2008gc} (EXO 1745-248), and \cite{Ozel:2008kb} (4U
      1608-52). \protect\label{fig:TOV_QMC}}
  \end{figure}

  The predicted values of mass and radius for QMC both with and
  without a phase transition to MIT bag model quark-matter are shown
  in Fig.~\ref{fig:TOV_QMC}. We note that the inclusion of a phase
  transition lowers the stellar mass for a star with a given central
  density, due to a further softening of the EoS.

\section{CONCLUSIONS}

  The efficiency with which QMC allows one to model hyperon
  contributions highlights the importance of baryon structure. The
  lack of a phase transition between hyperonic QMC and NJL
  quark-matter indicates the importance of both hyperon degrees of
  freedom and dynamic chiral symmetry breaking, both of which must be
  treated with care. A transition is possible when one or both of
  these factors is neglected, but the physics must be {\em removed} by
  hand.\par

  In order to achieve a Gibbs phase transition between a baryon phase
  and a quark-matter phase in which dynamical chiral symmetry breaking
  gives rise to constituent-quark masses, a moderately stiff baryon
  EoS is required, otherwise no such transition is possible. Such is
  the case for hyperonic QMC and NJL quark-matter.\par

  While the QMC model at Hartree level successfully reproduces many
  properties of finite nuclei and stellar objects, the softness of the
  hyperonic/hybrid EoS (as evidenced by the maximum stellar mass)
  %
  %
  is unable to fully account for the most massive pulsars observed
  to date. We note that further improvements such as calculations at
  Hartree--Fock level (of particular interest are the Fock terms
  corresponding to the $\pi$ meson) are currently in development.


\begin{theacknowledgments}
  This research was supported by the
  Australian Research Council. The author would like to thank
  A.~W.~Thomas for his guidance and support, as well as
  D.~B.~Leinweber and A.~G.~Williams for their helpful discussions.
\end{theacknowledgments}



\bibliographystyle{aipproc}   


\begin{thebibliography}{4}

  \bibitem{QMC2007}
    J.~Rikovska-Stone, P.~A.~M.~Guichon, H.~H.~Matevosyan, A.~W.~Thomas. 
    Nucl. Phys. A792:341-369, 2007 [doi:10.1016/j.nuclphysa.2007.05.011].

  \bibitem{QMC2008}
    P.~A.~M.~Guichon, A.~W.~Thomas, K.~Tsushima. 
    Nucl. Phys. A814:66-73, 2008 [doi:10.1016/j.nuclphysa.2008.10.001].

  \bibitem{Serot:1984ey}
    B.~D.~Serot, J.~D.~Walecka.
     Adv. Nucl. Phys.16:1-327, 1986. 

  \bibitem{JDC2009}
    J.~D.~Carroll, D.~B.~Leinweber, A.~G.~Williams, A.~W.~Thomas. 
    Phys. Rev. C79:045810, 2009 [doi:10.1103/PhysRevC.79.045810].

  \bibitem{JDCthesis}
    J.~D.~Carroll.
    [arXiv:1001.4318 [hep-ph]].
   
  \bibitem{Chodos:1974je}
    A.~Chodos, R.~L.~Jaffe, K.~Johnson, C.~B.~Thorn, V.~F.~Weisskopf.
    Phys. Rev.  D 9, 3471 (1974) [doi:10.1103/PhysRevD.9.3471].

  \bibitem{Nambu:1961tp}
    Y.~Nambu, G.~Jona-Lasinio.
    Phys. Rev. 122, 345 (1961) [doi:10.1103/PhysRev.122.345].

   \bibitem{Ozel:2006bv}
     F.~Ozel.
     Nature 441, 1115 (2006) [doi:10.1038/nature04858].

   \bibitem{Guver:2008gc}
     T.~Guver, F.~Ozel, A.~Cabrera-Lavers, P.~Wroblewski.
     arXiv:0811.3979 [astro-ph].

   \bibitem{Ozel:2008kb}
     F.~Ozel, T.~Guver, D.~Psaltis.
     Astrophys. J. 693, 1775 (2009) [doi:10.1088/0004-637X/693/2/1775].

\end{thebibliography}



\end{document}